# Spatial Scales of Population Synchrony generally increases as fluctuations propagate in a Two Species Ecosystem

Miguel Ángel Fernández-Grande[1*], Francisco Javier Cao-García[1,2**]

1. Departamento de Estructura de la Materia, Física Térmica y Electrónica, Universidad Complutense de Madrid, Parque de Ciencias, 1, 28040 Madrid, Spain.
2. Instituto Madrileño de Estudios Avanzados en Nanociencia, IMDEA Nanociencia, Calle Faraday, 9, 28049 Madrid, Spain.

* email: migufe03@ucm.es

** email: francao@ucm.es

## Abstract:

The spatial scale of population synchrony gives the characteristic distance at which the population fluctuations are correlated. Therefore, it gives also the characteristic size of the regions of simultaneous population depletion, or even extinction. Single-species previous results imply that the spatial scale of population synchrony is equal or greater (due to dispersion) than the spatial scale of synchrony of environmental fluctuations. Theoretical results on multispecies ecosystems points that interspecies interactions modify the spatial scale of population synchrony. In particular, recent results on two species ecosystems, for two competitors and for predator-prey, point that the spatial scale of population synchrony generally increases as the fluctuations propagates through the food web, i.e., the species more directly affected by environmental fluctuations presents the smaller spatial scale of population synchrony. Here, we found that this behaviour is generally true for a two species ecosystem. The exception to this behaviour are the particular cases where the population fluctuations of one of the species does not damp by its own, but requires a strong transfer of the fluctuation to the other species to be damped. These analytical results illustrate the importance of applying an ecosystem rather than a single-species perspective when developing sustainable harvestings or assessing the extinction risk of endangered species.

# I. Introduction

The spatial scales of population synchrony provide the distance at which the population fluctuations of different regions remain correlated. This information is relevant for different aspects such as the development of sustainable crops or the protection of endangered species (Persson, Van Leeuwen, and De Roos 2014; Strong and Frank 2010). In particular, the spatial scale of population synchrony gives the characteristic size of the regions with risk of regional extinction (M. Heino et al. 1997; Mikko Heino 1998a).

A main result in the beginning of the research in the spatial scales of population synchrony was when Moran, in Ref. (Moran 1953), showed that the synchrony between the size of two subpopulation is the same as the synchrony of the environmental fluctuations acting on them. He employed a simple lineal model without dispersion to this end. Later, in Ref. (Lande, Engen, and Sæther 1999) was proved that dispersion increases the spatial scale of population synchrony. These two results are for single species models. However, including additional interacting species in the model changes the spatial scales of population synchrony from those each species has separately without interaction (Bjørnstad, Ims, and Lambin 1999).

The influence of the inter-species interaction in the degree of spatial synchrony can also be seen in (Blasius, Huppert, and Stone 1999), where a three level trophic ecosystem is used to study periodic cycles in wildlife; in (Cazelles and Boudjema 2001), where the Moran effect is extended to complex non-linear dynamics; and in (Vasseur and Fox 2009), where it is shown that environmental fluctuations can stabilize food webs dynamics by increasing synchrony. The interaction between species may have different implications on spatial scales, i.e., depending on the type of interaction the spatial scales will increase or decrease, and the effect could be different for each one of the species, as was shown in (Jarillo et al. 2018) for two competing species, and in (Jarillo et al. 2020) for a predator and its prey. These two later works also showed that the strength of interspecific interaction may strongly modify the effect of harvesting on the spatial scales of population synchrony. They also showed that this two particular cases (two competitor and predator prey) share the common feature that generally the more perturbed species by the environmental fluctuations has the smaller spatial scale of population synchrony, pointing to an increase of the spatial scale of population synchrony as the perturbation propagates through a food web. Here, we address the questions on whether this statement is also valid for a general two species ecosystem, and we found it is valid and characterised the domain of validity.

The spatial scales of population synchrony in multispecies ecosystems have various applications such as the sustainable exploitation of the environment (Baskett, Micheli, and Levin 2007) or the protection of endangered species (Liebhold, Koenig, and Bjørnstad 2004). Also, they allow us to see some consequences of climate change, in (Both et al. 2004) the large scale of geographical variations indicates that climate change is advancing the lay of eggs by birds.

In section II we introduce a two species interacting model with dispersion and environmental stochasticity. In section III we derive and present the main result, generally the more perturbed species has the smaller spatial scale of population synchrony, explicitly solving the conditions for this result and their exceptions. Finally, in section IV we discuss the results and their applications and consequences as well as their limitations. Additionally, in Appendix A we give more details on the derivations, and the analytical expression of the spatial scales of population synchrony for a general two species ecosystem close to a stable equilibrium point and subject to small environmental fluctuations.

## II. General two interacting species ecosystem model

Here we introduce a mathematical model which describes the population dynamics of a general two species ecosystem with interspecific interaction.

### II.1. Deterministic model

In the first place, we set out a system of differential equations, which models the population dynamics between the two species at a single location. This model is local and deterministic (dispersion and environmental fluctuations are not included).

$$\begin{cases} \dfrac{dN_1}{dt} = R_1(N_1, N_2), \\ \dfrac{dN_2}{dt} = R_2(N_1, N_2), \end{cases} \quad (1)$$

where $N_i$ are the population size (or population density) of the specie $i$, and $R_i$ are smooth functions giving their temporal variations and they depend on the population size of the two species. The shape of these functions, $R_i$, defines the interaction dynamics of the ecosystem, i.e., whether the system is competitive, predator-prey, or has another kind of dynamic.

The functions $R_i$ can be non-linear, so the populations dynamics can be non-linear. Previous numerical simulations of non-linear models (Ripa and Ranta 2007; Engen and Lande 1996; Beddington and May 1977) have shown that linear approximations around equilibrium points presents the same dynamic than the non-linear model, for small fluctuations. This enables us to study the system close to the stable equilibrium.

The equilibrium points $(N_1^{eq}, N_2^{eq})$ of the system verify the condition $R_i(N_1^{eq}, N_2^{eq}) = 0$, for $i = 1, 2$. We express the population size around the equilibrium as $N_i = N_i^{eq}(1 + \epsilon_i)$ where $\epsilon_i$ are the relative populations fluctuations. Neglecting second or higher order of these fluctuations the dynamics of the model may be rewritten as

$$\frac{d}{dt}\begin{pmatrix} \epsilon_1 \\ \epsilon_2 \end{pmatrix} = -\begin{pmatrix} \gamma_{11} & \gamma_{12} \\ \gamma_{21} & \gamma_{22} \end{pmatrix}\begin{pmatrix} \epsilon_1 \\ \epsilon_2 \end{pmatrix} = -\Gamma \begin{pmatrix} \epsilon_1 \\ \epsilon_2 \end{pmatrix}. \tag{2}$$

where $\gamma_{ij} = -\frac{\partial R_i(N_1^{eq}, N_2^{eq})}{\partial N_j}$. The equilibrium characterised by the $\Gamma$ matrix will be stable if the eigenvalues of the matrix have positive real part, or equivalently the trace, $\tau$, and the determinant, $\Delta$, are positive, i.e.,

$$\begin{aligned} \Delta &= \gamma_{11}\gamma_{22} - \gamma_{12}\gamma_{21} > 0, \\ \tau &= \gamma_{11} + \gamma_{22} > 0. \end{aligned} \tag{3}$$

### II.2. Spatial stochastic model

In the second place, we introduce environmental fluctuations and dispersion. We begin with the environmental fluctuations. To do this we add the term $N_i \sigma_{ei} dB_i(t)$ to each equation in (1):

$$\begin{cases} \frac{dN_1}{dt} = R_1(N_1, N_2) + N_1 \sigma_{e1} dB_1(t), \\ \frac{dN_2}{dt} = R_2(N_1, N_2) + N_2 \sigma_{e2} dB_2(t), \end{cases} \tag{4}$$

where $\sigma_{ei}^2$ is the environmental variance acting on species $i$ and $dB_i(t)$ is the infinitesimal increment of a standard Brownian motion with $E[dB_i(t)] = 0$ y $E[dB_i(t)^2] = dt$. The two noise terms may in general be correlated, $E[dB_1(t)dB_2(t)] = \rho_{12} dt$.

Until here we have just considered the population dynamics at a single location, from this point on we are going to include spatial processes. Thus, the population fluctuations of each species at position $x = (x_1, x_2) \in \mathbb{R}^2$ and time $t$, $\vec{\epsilon}(x, t)$, can be described by the matrix equation

$$d\vec{\epsilon}(x,t) = -(\Gamma + M)\vec{\epsilon}(x,t)dt + d\vec{A}(x,t), \tag{5}$$

where the vector $d\vec{A}(x,t) = \Sigma_e d\vec{B}(x,t)$ contains the environmental noises. In the previous equation $\Sigma_e$ represents a diagonal matrix with elements $\sigma_{ei}$, while $d\vec{B}$ are increments of correlated standard Brownian motions with $E[dB_i(x,t)dB_j(x+y,t)] = \rho_{ij}(y)dt$ with $\rho_{ii}(0) = 1$ and $|\rho_{12}(0)| \leq 1$. Moreover $E[d\vec{A}(x)d\vec{A}(x+y,t)] = P(y)dt$. The elements of the covariance matrix $P(y)$ are $\sigma_{ei}\sigma_{ej}\rho_{ij}(y)$. This way, $P_{ii}(0) = \sigma_{ei}^2$ is the environmental variance and $\rho_{ii}(y)$ is the spatial autocorrelation of the noise acting on species $i$, while $\rho_{12}(y)$ is the spatial autocorrelation between the noise terms acting on the species at distance $y$. Finally, we can express the dispersal contribution to the dynamics as

$$M \equiv \begin{pmatrix} m_1 - m_1(f_1 *.) & 0 \\ 0 & m_2 - m_2(f_2 *.) \end{pmatrix}, \tag{6}$$

where $(f_i *.)$ is the convolution operator, whose linear action on an arbitrary function $g(x,t)$ is given by $(f_i * g)(x,t) = \int_{-\infty}^{\infty} g(x-y,t)f_i(y)dy$.

### II.3. Spatial scales of population synchrony

From the population fluctuations we can define the cross-covariance elements $c_{ij}(y) \equiv E[\epsilon_i(x,t_0)\epsilon_j(x+y,t_0)]$. We introduce the spatial covariance matrix of the population density fluctuations as

$$C(y) = \begin{pmatrix} c_{11}(y) & c_{12}(y) \\ c_{21}(y) & c_{22}(y) \end{pmatrix}. \tag{7}$$

$c_{ii}(0)$ provides the change of the relative population fluctuations of species $i$, $c_{ii}(y)$ relies on the size and the synchrony of the relative population fluctuations of each species at a position y. The non-diagonal terms $c_{12}(y)$ and $c_{21}(y)$ satisfy that $c_{21}(y) = c_{12}(-y)$, and they rely on the size and the synchrony of the relative population fluctuations of species 1 and 2 separated a distance $y$.

Because of the stationarity of the process, the cross-covariance elements, $c_{ij}(y)$ are time independent, so $c_{ij}(y) = E[\epsilon_i(x,t_0)\epsilon_j(x+y,t_0)] = E[\epsilon_i(x,t_0+dt)\epsilon_j(x+y,t_0+dt)]$. Replacing in the $c_{ij}(y)$ time independent expression the differential variations in the population fluctuations in a different time step $dt$, Eq. (5), and neglecting second order terms on $dt$, we gain the equation

$$C(y)(\Gamma' + M') + (\Gamma + M)C(y) = P(y). \tag{8}$$

We can see an analogous expression to the latter equation in Eq. (3) of Ref. (Lande, Engen, and Sæther 1999) where the single species case is analysed. The Eq. (8) here is the multispecies generalization of that single species equation.

Then, we can use (9) to calculate the integral of the covariances over the hold space,

$$I_{ij} \equiv \int c_{ij}(y) dy. \tag{9}$$

Due to the symmetry of the problem, we can study the spatial scales of population synchrony along a given direction chosen as the direction of the first axis without loss of generality. For positive covariances, we can define the spatial scales of population synchrony in this direction as

$$l_{ij}^2 \equiv \frac{1}{I_{ij}} \int y_1^2 c_{ij}(y) dy = -\frac{1}{4\pi^2} \frac{\frac{\partial^2 \hat{c}_{ij}(k)}{\partial k_1^2}}{\hat{c}_{ij}(k)} \bigg|_{k=0}, \tag{10}$$

with $\hat{c}_{ij}(k) \equiv \int c_{ij}(x) \exp(-2\pi i k x) \, dx$ the Fourier spatial transform of $c_{ij}(x)$. The later or the two expressions in Eq. (10) is particularly convenient for analytical computations. The Fourier transformation of Eq. (8) converts the dispersal convolutions (contained in $M$) into products of the autocovariances $\hat{c}_{ii}(k)$ and the dispersal functions $\hat{f}_i(k)$ in Fourier space. This makes that the solution of Eq. (8) becomes an algebraic problem due to its linearity in the covariances. The general results for $l_{ij}^2$ are given in Appendix A (Jarillo et al. 2018).

# III. Results: Spatial Scales of Population Synchrony generally increases as fluctuations propagate in a Two Species Ecosystem

Here we address whether the spatial scale of population synchrony is smaller for the species more affected by the environmental fluctuations. We use the general results for the spatial scales of population synchrony have been obtained following the procedure in the previous section. (See Appendix A for the full analytical expressions resulting.)

We focus first on the case where the environmental noise of the second specie is bigger than the environmental noise of the first specie ($\sigma_{e2} \gg \sigma_{e1}$), which implies for the differences of the square of spatial scales of population synchrony

$$(l_{11}^2 - l_{22}^2) \xrightarrow{\sigma_{e2} \gg \sigma_{e1}} \frac{\gamma_{11}(m_1 l_{m1}^2 + m_2 l_{m2}^2) + \tau\, m_1 l_{m1}^2}{\gamma_{11}^2 + \Delta}. \tag{11}$$

Defining, the rate of migration capacities as $d_{21} = (m_2 l_{m2}^2)/(m_1 l_{m1}^2)$, Eq. (11) can be rewritten as

$$(l_{11}^2 - l_{22}^2) \xrightarrow{\sigma_{e2} \gg \sigma_{e1}} \frac{m_1 l_{m1}^2}{\gamma_{11}^2 + \Delta} \left[\gamma_{11}(2 + d_{21}) + \gamma_{22}\right]. \tag{12}$$

The first factor is definite positive. Thus, the second factor, the bracket, gives the sign, determining which of the spatial scales of synchrony is larger when the environmental noise of the second species dominates. Additionally, the rate of migration capacities is always positive, $d_{21} > 0$.

When the fluctuations are damped by its own we have that both of the diagonal damping rates are positive, $\gamma_{11} > 0$ and $\gamma_{22} > 0$, which implies that the smaller spatial scale of population synchrony is that of the species 2, i.e., the more affected by the environmental noise. However, stability requires a less restrictive condition, that the trace is positive, $\tau = \gamma_{11} + \gamma_{22} > 0$, allowing negative values of one of the two diagonal damping rates. The stability condition implies $\gamma_{22} > -\gamma_{11}$. Therefore, within the stability region we have two possible regimes

$$(l_{11}^2 - l_{22}^2) \xrightarrow{\sigma_{e2} \gg \sigma_{e1}} \begin{cases} > 0, & \text{if} \quad \gamma_{22}/\gamma_{11} > -(2 + d_{21}), \\ < 0, & \text{if} \quad -(2 + d_{21}) > \gamma_{22}/\gamma_{11} > -1. \end{cases} \tag{13}$$

The regions in the $(\gamma_{11}, \gamma_{22})$ plane of these two regimes (for $\gamma_{11} > 0$) are depicted in Panel A of Fig. 1. Fig. 1 also shows the differences of spatial scales of populations synchrony for three examples, two corresponding to the regime where $(l_{11}^2 - l_{22}^2)|_{\sigma_{e2} \gg \sigma_{e1}} > 0$, and another one to the other regime. Fig. 2 also shows in a color plot the sign and magnitude of the spatial scale of population synchrony as a function of the environmental noise ratio $\sigma_{e2}/\sigma_{e1}$ and of the diagonal damping rate of the second species $\gamma_{22}$. In the examples of the previous plots consider and compare the uncorrelated case $[\rho_{12}(y) = 0]$ and the completely correlated case $[\rho_{12}(y) = \rho_{11}(y) = \rho_{22}(y)]$ showing that both have the same asymptotic behaviour when the environmental noise of the second species dominates $\sigma_{e2} \gg \sigma_{e1}$. This is also true for the opposite limit, when it is the environmental noise of the first species which dominates $\sigma_{e1} \gg \sigma_{e2}$.

The results for the opposite limit, $\sigma_{e1} \gg \sigma_{e2}$, are easily obtained just switching indices

$$(l_{22}^2 - l_{11}^2) \xrightarrow{\sigma_{e1} \gg \sigma_{e2}} \frac{m_2 l_{m2}^2}{\gamma_{22}^2 + \Delta} [\gamma_{22}(2 + d_{12}) + \gamma_{11}]. \tag{14}$$

The stability condition implies $\gamma_{11} > -\gamma_{22}$. Therefore, within the stability region we have two possible regimes

$$(l_{22}^2 - l_{11}^2) \xrightarrow{\sigma_{e1} \gg \sigma_{e2}} \begin{cases} > 0, & \text{if} \quad \gamma_{11}/\gamma_{22} > -(2 + d_{12}), \\ < 0, & \text{if} \quad -(2 + d_{12}) > \gamma_{11}/\gamma_{22} > -1, \end{cases} \tag{15}$$

where $d_{12} = 1/d_{21} = (m_1 l_{m1}^2)/(m_2 l_{m2}^2)$. The regions of these two regimes (for $\gamma_{11} > 0$) are also depicted in the $(\gamma_{11}, \gamma_{22})$ plane in Panel A of Fig. 1. Therefore, Panel A of Fig. 1 shows the three regions resulting from the two different regimes in the two limits (first or second species environmental noises dominates). [The condition $\gamma_{11}/\gamma_{22} > -(2 + d_{12})$ can also be expressed as $\gamma_{22}/\gamma_{11} > -1/(2 + d_{12})$]

In summary, the spatial scale of population synchrony is smaller for the species more affected by environmental noise provided the following condition is satisfied

$$\frac{\gamma_{22}}{\gamma_{11}} > \max\left\{-(2 + d_{21}), -\frac{1}{(2 + d_{12})}\right\}. \tag{16}$$

This condition is stronger than the stability condition, $\gamma_{22}/\gamma_{11} > -1$, but it is fulfilled in particular in the cases where both diagonal damping rates are positive, $\gamma_{11} > 0$ and $\gamma_{22} > 0$.

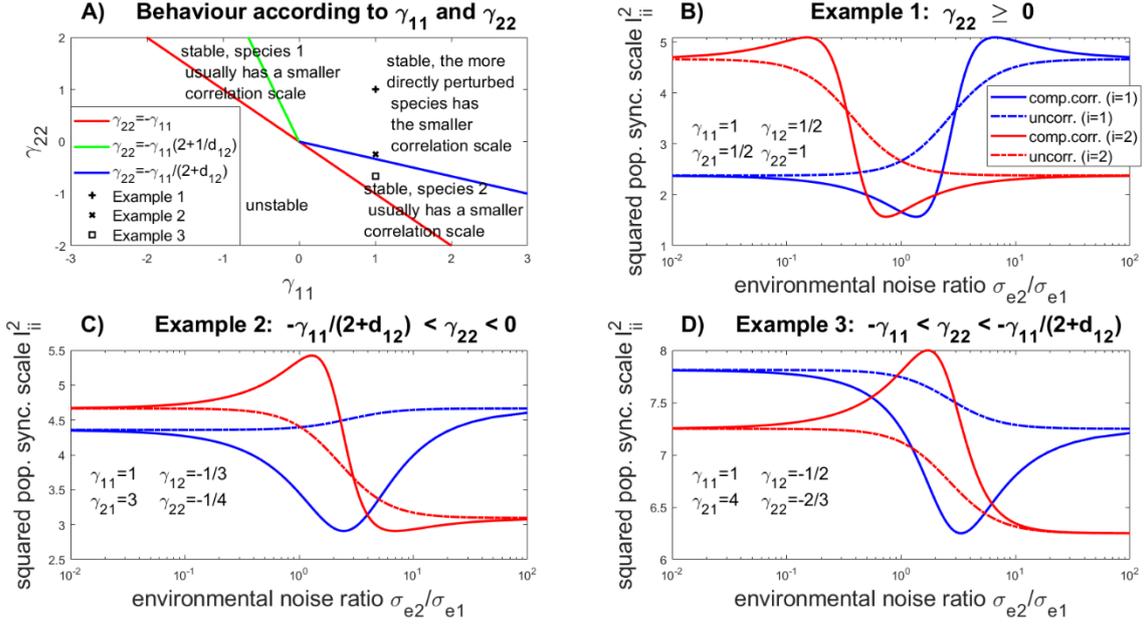

**Figure 1. Difference of squared spatial scales of population synchrony $l_{11}^2 - l_{22}^2$ of the two species as a function of their diagonal damping rates $\gamma_{11}$ and $\gamma_{22}$. A)** The $(\gamma_{11}, \gamma_{22})$ plane presents three subregions within the stable region, for the asymptotic behaviour of the difference of squared spatial scales of population synchrony $l_{11}^2 - l_{22}^2$ of the two species. In the first main stable region (upper-right of this panel) the more intensively perturbed species by environmental fluctuations has the smaller spatial scale of population synchrony. (Panels B and C represent examples of this case, examples 1 and 2.) In the other two stable regions (upper-left and down-right of this panel) one of the spatial scales of population synchrony tends to be always the smaller one, this regimes require one of the diagonal damping rates, $\gamma_{11}$ or $\gamma_{22}$, to be negative. (Panel D represents an example of this latter case, example 3.) **B)** In Example 1 ($\gamma_{22} \geq 0$) we can see that the more directly perturbed species has the smaller spatial scale of population synchrony, independently on whether the environmental noises on the species are completely correlated (solid lines, $\rho_{11}(y) = \rho_{22}(y) = \rho_{12}(y)$) or uncorrelated (dashed lines, $\rho_{12}(y) = 0$). The independence of the results from the correlation of the noises was expected due to independence of the result in Eq. (11). **C)** In Example 2 $\left(-\frac{\gamma_{11}}{2+d_{12}} < \gamma_{22} < 0\right)$ we have the same behaviour than in the previous example. The difference is that here one of the diagonal elements of the Γ matrix is negative. This shows illustrates that this asymptotic behaviour can also occur with one negative diagonal damping rate. **D)** In Example 3 $\left(-\gamma_{11} < \gamma_{22} < -\frac{\gamma_{11}}{2+d_{12}}\right)$ we observe that one of the species (the first one) always have the bigger correlation scales when the noise affecting one of the species is bigger than the noise affecting the other one, consistently with the regions represented in Panel A of this figure. All panels of this figure represent the case $m_1 l_{m1}^2 = m_2 l_{m2}^2 = 1$, $l_{e11}^2 = l_{e22}^2 = l_{e12}^2 = 1$ and $\rho_{11}(y) = \rho_{22}(y)$. Therefore, for the rate of diffusion capacities we have $d_{12} = \frac{m_1 l_{m1}^2}{m_2 l_{m2}^2} = \frac{1}{d_{21}} = 1$.

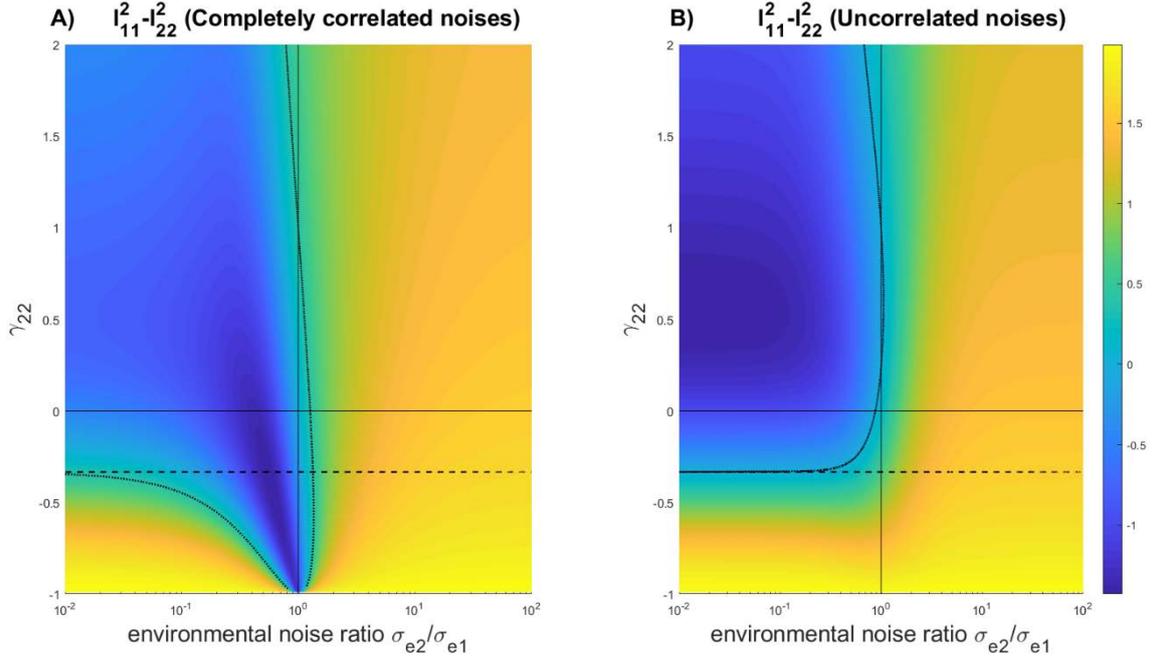

**Figure 2. Difference of squared spatial scales of population synchrony $l_{11}^2 - l_{22}^2$ as a function of the diagonal damping rate of the second species $\gamma_{22}$ and of the ratio of environmental noises $\sigma_{e2}/\sigma_{e1}$.** The other three elements of the $\Gamma$ matrix are fixed to: $\gamma_{11} = 1$, $\gamma_{12} = -1$ and $\gamma_{21} = 1$. The dotted line indicates when both spatial scales of correlation are equal. The dashed line separates the two different regimes $\left(\gamma_{22} = -\frac{\gamma_{11}}{2+d_{12}}\right)$ described in Fig. 1 and in Section III of the text. Above the dashed line the more directly perturbed species has the smaller spatial scale of population synchrony, below the dashed line is always species 2 that has the smaller spatial scale of population synchrony. Panel A shows the completely correlated noise case $\left(\rho_{11}(y) = \rho_{22}(y) = \rho_{12}(y)\right)$ and Panel B the uncorrelated noise case $(\rho_{12}(y) = 0)$. Both presents the same asymptotic behaviour. The independence of the results from the correlation of the noises was expected due to independence of the result in Eq. (11). All panels of this figure have the following values of the other parameters $m_1 l_{m1}^2 = m_2 l_{m2}^2 = 1$, $l_{e11}^2 = l_{e22}^2 = l_{e12}^2 = 1$ and $\rho_{11}(y) = \rho_{22}(y)$.

## IV. Discussion

Previous research on single species models have shown that the spatial correlation of the environmental noise and the dispersal of the species determines the spatial scale of population synchrony (Moran 1953; Lande, Engen, and Sæther 1999). These phenomena causes variations in the geographical distribution of the species (Levin 1992). Here, we show how these intraspecies effects are modulated by the interactions with other species. Specifically, we focus on finding out which is the species with the higher spatial scale of population synchrony when the ratio of environmental noise is big or small, i.e. when one of the environmental noises dominates.

We found that the criteria which determines the larger spatial scale of population synchrony, depends on the diagonal damping rates and on the dispersal rates and dispersal lengths. These characteristics of the results were to be expected in view of previous results (Engen, Lande, and Sæther 2002), where the connection between dispersal and the spatial scale of population synchrony is proved. One additional relevant aspect of the results is its independence from the correlation between the environmental noises for each species. This independence may seem surprising, but it is important to recall that this independence only occurs in the limit when one of the environmental noises dominates. Therefore the effects of the other environmental noise becomes negligible, including its correlation with the dominant environmental noise. Results outside this limit are not independent of the correlation between environmental noises, as can be seen in Panels B, C and D of Fig. 1.

Several theoretical studies (M. Heino et al. 1997; Mikko Heino 1998b; Earn, Levin, and Rohani 2000; Ripa and Ranta 2007; Engen 2007) have shown the link between the spatial scale of population synchrony and the size of the region suffering local extinction. Consequently, knowing the relative magnitudes of the spatial scales of population synchrony can be decisive when setting up strategies for a sustainable management of exploited species, e.g., deciding fishing quotas or harvesting strategies. Although the achieved result could seem scant general, since we have just considered the limits situations where one of the noises dominates, this situation is very common in the nature and in the ecosystems models (McCann 2012; Terborgh 2015). For instance, a simple example of this situation are the predator-prey ecosystems with bottom-up regulation, where the environmental noises affecting the prey are bigger than the ones affecting the predator (Jarillo et al. 2020). In top-down ecosystems we have the opposite situation.

Finally, our model only considers two species, adding the interactions with the other species living in the ecosystem to the environmental noise. A more realistic analysis should deal with the dynamics of all the species in the ecosystem, however it seems difficult to obtain general analytical results in these general multispecies models. The next step between an entirely realistic analysis and this one, could be the incorporation of a third species. This three species model could let us study different situations such as the coexistence of two different species with a common predator or the interactions between species of a simple three level trophic web.

# Acknowledgements

This work was supported by the European Economic Area (EEA) Grants UCM-EEA-ABEL-02-2009 and 005-ABEL-CM2014A under the NILS – Science and Sustainability Programme, by the SUSTAIN project of the Research Council of Norway and by the Research Council of Norway (SFF-III 223257/F50). FJCG acknowledges financial support through grants FIS2010-17440 and FIS2006-05895 of Ministerio de Ciencia e Innovación (Spain), FIS2015-67745-R (MINECO/FEDER) of Ministerio de Economía y Competitividad (Spain) and Fondo Europeo de Desarrollo Regional (FEDER, EU), RTI2018-095802-B-I00 of Ministerio de Ciencia, Innovación y Universidades (Spain), GR35/14-920911, GR35/10-A-920911 and GR58/08-920911 of Universidad Complutense de Madrid and Banco Santander (Spain), 817578 TRIATLAS of Horizon 2020 Programme (EU).

**Appendix A: Analytical expressions for the spatial scales of population synchrony in a general two species ecosystem.**

The population fluctuations for each species in a given position and time can be expressed by the following matrix equation:

$$d\vec{\epsilon}(x,t) = -(\Gamma + M)\vec{\epsilon}(x,t)dt + d\vec{A}(x,t), \quad (A1)$$

where $\vec{\epsilon}$ symbolises the fluctuations, $\Gamma$ is the equilibrium matrix from the deterministic model, $M$ gives the dispersal contribution and $d\vec{A}$ contains the environmental noises.

Let $C(y)$ be the spatial covariance matrix of the population density fluctuations, with $c_{ij}(y) = E[\epsilon_i(x,t_0)\epsilon_j(x+y,t_0)]$. Due to the stationarity of the process, replacing $c_{ij}(y)$ in two different time steps in Equation A1 and neglecting second order terms on $dt$ we obtain:

$$C(y)(\Gamma' + M') + (\Gamma + M)C(y) = P(y), \quad (A2)$$

where $'$ denotes transposition and $P(y)$ is the covariance matrix with elements $\sigma_{ei}\sigma_{ej}\rho_{ij}(y)$. Using Equation A2 we can compute the integral of the covariance over the hold space

$$I_{ij} = \int c_{ij}(y)dy.$$

As a consequence of the symmetry of the problem we can study the spatial scales along a given direction as the direction of the first axis without loss of generality. Then, for positive covariances we can define the spatial scales of population synchrony in this direction as

$$l_{ij}^2 = \frac{1}{I_{ij}} \int y^2 c_{ij}(y) dy = -\frac{1}{4\pi^2} \frac{\frac{\partial^2 \widehat{c_{ij}}(k)}{\partial k^2}}{\widehat{c_{ij}}(k)}\Bigg|_{k=0} \qquad (A3)$$

where $\widehat{c_{ij}}(k) = \int c_{ij}(x) \exp(-2\pi i k x) \, dx$ is the Fourier spatial transform of $c_{ij}(x)$. In order to obtain the expressions for the spatial scales of population synchrony we will use the definition in terms of the Fourier spatial transform. When we apply the Fourier spatial transform to Equation A2, the transformed elements of $C$ form an algebraic equation system that we can solve. Then we just derive the $\widehat{c_{ij}}(k)$ and we gain $l_{ij}^2$.

The expressions for the spatial scales of population synchrony read

$$l_{11}^2 = \frac{\gamma_{12}^2 J_{e22}(l_{e22}^2 - l_{e11}^2) - 2\gamma_{12}\gamma_{22}J_{e12}\left(l_{e12}^2 - l_{e11}^2 - \frac{m_2 l_{m2}^2}{\gamma_{22}}\right) - J_{e11}\gamma_{22}\left(\gamma_{11}\left(\frac{m_1 l_{m1}^2}{\gamma_{11}} + \frac{m_2 l_{m2}^2}{\gamma_{22}}\right) + 2\gamma_{22}\frac{m_2 l_{m2}^2}{\gamma_{22}}\right)}{\gamma_{12}^2 J_{e22} - 2\gamma_{12}\gamma_{22}J_{e12} + J_{e11}(\gamma_{22}(\gamma_{11} + \gamma_{22}) - \gamma_{12}\gamma_{21})}$$

$$+ l_{e11}^2 + \frac{m_1 l_{m1}^2 + m_2 l_{m2}^2}{\gamma_{11} + \gamma_{22}} + \frac{\gamma_{11}\gamma_{22}}{\gamma_{11}\gamma_{22} - \gamma_{12}\gamma_{21}}\left(\frac{m_1 l_{m1}^2}{\gamma_{11}} + \frac{m_2 l_{m2}^2}{\gamma_{22}}\right), \qquad (A4)$$

$$l_{22}^2 = \frac{\gamma_{21}^2 J_{e11}(l_{e11}^2 - l_{e22}^2) - 2\gamma_{21}\gamma_{11}J_{e12}\left(l_{e12}^2 - l_{e22}^2 - \frac{m_1 l_{m1}^2}{\gamma_{11}}\right) - J_{e22}\gamma_{11}\left(\gamma_{22}\left(\frac{m_1 l_{m1}^2}{\gamma_{11}} + \frac{m_2 l_{m2}^2}{\gamma_{22}}\right) + 2\gamma_{11}\frac{m_1 l_{m1}^2}{\gamma_{11}}\right)}{\gamma_{21}^2 J_{e11} - 2\gamma_{21}\gamma_{11}J_{e12} + J_{e22}(\gamma_{11}(\gamma_{11} + \gamma_{22}) - \gamma_{12}\gamma_{21})}$$

$$+ l_{e22}^2 + \frac{m_1 l_{m1}^2 + m_2 l_{m2}^2}{\gamma_{11} + \gamma_{22}} + \frac{\gamma_{11}\gamma_{22}}{\gamma_{11}\gamma_{22} - \gamma_{12}\gamma_{21}}\left(\frac{m_1 l_{m1}^2}{\gamma_{11}} + \frac{m_2 l_{m2}^2}{\gamma_{22}}\right), \qquad (A5)$$

where

$$J_{eij} = \sigma_{ei}\sigma_{ej} \int \rho_{ij}(y) dy,$$

$$l_{eij}^2 = \frac{\sigma_{ei}\sigma_{ej}}{J_{eij}} \int y^2 \rho_{ij}(y) dy = \frac{\int y^2 \rho_{ij}(y) dy}{\int \rho_{ij}(y) dy},$$

$$l_{mi}^2 = \int y^2 f_i(y) dy.$$

However, we are interested in the difference between the spatial scales of population synchrony of each species. The expression for the difference is

$$l_{11}^2 - l_{22}^2 = l_{e11}^2 - l_{e22}^2 +$$

$$+ \frac{a_{1111}J_{e11}^2 + a_{2222}J_{e22}^2 + a_{1212}J_{e12}^2 + a_{1122}J_{e11}J_{e22} + a_{1211}J_{e12}J_{e11} + a_{1222}J_{e12}J_{e22}}{b_{1111}J_{e11}^2 + b_{2222}J_{e22}^2 + b_{1212}J_{e12}^2 + b_{1122}J_{e11}J_{e22} + b_{1211}J_{e12}J_{e11} + b_{1222}J_{e12}J_{e22}}, \qquad (A6)$$

where

$$a_{1111} = -\gamma_{21}^2\left(\gamma_{22}\left(\gamma_{11}\left(\frac{m_1 l_{m1}^2}{\gamma_{11}} + \frac{m_2 l_{m2}^2}{\gamma_{22}}\right) + 2\gamma_{22}\frac{m_2 l_{m2}^2}{\gamma_{22}}\right) + (l_{e11}^2 - l_{e22}^2)(\gamma_{22}(\gamma_{11} + \gamma_{22}) - \gamma_{12}\gamma_{21})\right),$$

$$a_{2222} = \gamma_{12}^2\left(\gamma_{11}\left(\gamma_{22}\left(\frac{m_1 l_{m1}^2}{\gamma_{11}} + \frac{m_2 l_{m2}^2}{\gamma_{22}}\right) + 2\gamma_{11}\frac{m_1 l_{m1}^2}{\gamma_{11}}\right) + (l_{e22}^2 - l_{e11}^2)(\gamma_{11}(\gamma_{11} + \gamma_{22}) - \gamma_{12}\gamma_{21})\right),$$

$$a_{1212} = 4\gamma_{11}\gamma_{22}\gamma_{12}\gamma_{21}\left(l_{e22}^2 - l_{e11}^2 + \frac{m_1 l_{m1}^2}{\gamma_{11}} - \frac{m_2 l_{m2}^2}{\gamma_{22}}\right),$$

$$a_{1122} = 2\gamma_{12}^2\gamma_{21}^2(l_{e22}^2 - l_{e11}^2) + (\gamma_{22}m_1 l_{m1}^2 + \gamma_{11}m_2 l_{m2}^2)(\gamma_{22}^2 - \gamma_{11}^2) + 2\gamma_{11}\gamma_{22}(\gamma_{11} + \gamma_{22})(m_1 l_{m1}^2 - m_2 l_{m2}^2)$$
$$+ 2\gamma_{12}\gamma_{21}(\gamma_{22}m_2 l_{m2}^2 - \gamma_{11}m_1 l_{m1}^2),$$

$$a_{1211} = 2\gamma_{21}\left(m_1 l_{m1}^2(\gamma_{12}\gamma_{21} - \gamma_{22}^2) + m_2 l_{m2}^2(\gamma_{12}\gamma_{21} + \gamma_{11}(\gamma_{11} + 2\gamma_{22})) + 2\gamma_{22}\gamma_{12}^2 l_{e11}^2\right.$$
$$\left. + l_{e12}^2(\gamma_{11}\gamma_{22} - \gamma_{12}\gamma_{21})(\gamma_{11} + \gamma_{22}) - l_{e22}^2(\gamma_{11}\gamma_{22}(\gamma_{11} + \gamma_{22}) - \gamma_{12}\gamma_{21}(\gamma_{11} - \gamma_{22}))\right),$$

$$a_{1222} = -2\gamma_{12}(m_2 l_{m2}^2(\gamma_{12}\gamma_{21} - \gamma_{11}^2) + m_1(\gamma_{12}\gamma_{21} + \gamma_{22}(2\gamma_{11} + \gamma_{22}))$$
$$+ 2\gamma_{12}\gamma_{21}\gamma_{11}l_{e22}^2 + l_{e12}^2(\gamma_{11}\gamma_{22} - \gamma_{12}\gamma_{21})(\gamma_{11} + \gamma_{22}) + l_{e11}^2(\gamma_{12}\gamma_{21}(\gamma_{22} - \gamma_{11})$$
$$- \gamma_{11}\gamma_{22}(\gamma_{11} + \gamma_{22}),$$

$$b_{1111} = \gamma_{21}^2(\gamma_{22}(\gamma_{11} + \gamma_{22}) - \gamma_{12}\gamma_{21}),$$

$$b_{2222} = \gamma_{12}^2(\gamma_{11}(\gamma_{11} + \gamma_{22}) - \gamma_{12}\gamma_{21}),$$

$$b_{1212} = 4\gamma_{11}\gamma_{12}\gamma_{21}\gamma_{22},$$

$$b_{1122} = \left((\gamma_{11} + \gamma_{22})(\gamma_{12}^2\gamma_{22} + \gamma_{21}^2\gamma_{11}) - \gamma_{12}\gamma_{21}(\gamma_{12}^2 + \gamma_{21}^2)\right),$$

$$b_{1211} = -2\left(\gamma_{12}\gamma_{21}^2(\gamma_{22} - \gamma_{11}) + \gamma_{22}(\gamma_{11} + \gamma_{22})\right),$$

$$b_{1222} = -2\left(\gamma_{21}\gamma_{12}^2(\gamma_{11} - \gamma_{22}) + \gamma_{11}(\gamma_{11} + \gamma_{22})\right).$$

Regarding the environmental noise ratio, we will focus on two specific situations: when the environmental noise of the second specie is bigger than the environmental noise of the first specie ($\sigma_{e2} \gg \sigma_{e1} \to J_{e22} \gg J_{e12}, J_{e11}$) and vice versa ($\sigma_{e1} \gg \sigma_{e2} \to J_{e11} \gg J_{e12}, J_{e22}$).
Applying these two limits to Eq A3, we obtain the following expressions:

$$\sigma_{e2} \gg \sigma_{e1} \to l_{11}^2 - l_{22}^2 \approx l_{e11}^2 - l_{e22}^2 + \frac{a_{2222}}{b_{2222}} = \frac{\gamma_{11}(m_1 l_{m1}^2 + m_2 l_{m2}^2) + \tau m_1 l_{m1}^2}{\gamma_{11}^2 + \Delta} \quad (A7)$$

$$\sigma_{e1} \gg \sigma_{e2} \to l_{22}^2 - l_{11}^2 \approx l_{e22}^2 - l_{e11}^2 - \frac{a_{1111}}{b_{1111}} = \frac{\gamma_{22}(m_1 l_{m1}^2 + m_2 l_{m2}^2) + \tau m_2 l_{m2}^2}{\gamma_{22}^2 + \Delta}, \quad (A8)$$

where, $\tau = \gamma_{11} + \gamma_{22}$ and $\Delta = \gamma_{11}\gamma_{22} - \gamma_{12}\gamma_{21}$ are the trace and the determinant of the $\Gamma$ matrix from the deterministic model.